\documentclass{aastex63}

\shorttitle{Eight Year unWISE Coadds}
\shortauthors{Meisner et al.}

\graphicspath{{./}{figures/}}

\begin{document}

\title{Eight-year Full-depth unWISE Coadds}

\correspondingauthor{Aaron M. Meisner}
\email{aaron.meisner@noirlab.edu}

\author[0000-0002-1125-7384]{Aaron M. Meisner}
\affiliation{NSF's National Optical-Infrared Astronomy Research Laboratory, 950 N. Cherry Ave., Tucson, AZ 85719, USA}

\author[0000-0002-1172-0754]{Dustin Lang}
\affiliation{Perimeter Institute for Theoretical Physics, Waterloo, ON N2L 2Y5, Canada}

\author[0000-0002-3569-7421]{Edward F. Schlafly}
\affiliation{Lawrence Livermore National Laboratory 7000 East Avenue, Livermore, CA 94550, USA}

\author[0000-0002-5042-5088]{David J. Schlegel}
\affiliation{Physics Division, Lawrence Berkeley National Laboratory, 1 Cyclotron Road, Berkeley, CA 94720, USA}

\begin{abstract}

We present deep, full-sky maps built from \textit{Wide-field Infrared Survey Explorer} (\textit{WISE}) and NEOWISE exposures spanning the 2010 January - 2020 December time period. These coadds, which incorporate roughly 8 years of W1 (3.4~$\mu$m) and W2 (4.6~$\mu$m) imaging, are the deepest ever full-sky maps at wavelengths of 3-5~$\mu$m. Photometry based on these coadds will be a component of DESI Legacy Imaging Surveys DR10.

\end{abstract}


\section{Introduction} \label{sec:intro}

The \textit{Wide-field Infrared Survey Explorer} \citep[\textit{WISE};][]{wright10} provides a uniquely sensitive full-sky view of the universe at mid-infrared wavelengths. From early 2010 to early 2011, \textit{WISE} surveyed the entire sky 2+ times in both its W1 and W2 channels. \textit{WISE} was then placed into hibernation for nearly three years. In late 2013, \textit{WISE} was reactivated via the NEOWISE mission \citep{neowise,neowiser}. As of mid-2021, the post-reactivation NEOWISE mission has supplied and publicly released an additional 14 full-sky passes of W1/W2 imaging acquired during 7 years of continuous operation. NEOWISE is primarily driven by asteroid characterization science, and so does not publish coadded data products. unWISE \citep{lang14, lang14b} has therefore embarked on a sustained multi-year effort to maximize the combined value of \textit{WISE} and NEOWISE imaging for Galactic and extragalactic science, by creating a variety of deep, coadded full-sky maps and catalogs \citep{fulldepth_neo1, fulldepth_neo2, tr_neo2, fulldepth_neo3,tr_neo3,neo4_coadds, unwise_catalog, fulldepth_neo5, fulldepth_neo6}. unWISE has largely operated within the context of the DESI Legacy Imaging Surveys \citep{dey_overview}, as \textit{WISE} mid-infrared fluxes benefit DESI target selection \citep{desi_ts_prelim_bgs,desi_ts_prelim_elg,desi_ts_prelim_lrg,desi_ts_prelim_mw,desi_ts_prelim_qso}.

\section{Results}

We downloaded the full set of W1 and W2 single-exposure (``L1b'') images that were publicly released by NEOWISE on 2021 March 24. These images span from 2019 December 13 to 2020 December 13 (UTC). In combination with our existing downloads of all previous \textit{WISE} and NEOWISE L1b data releases, we thus have local access to copies of all publicly available single-exposure W1 and W2 images acquired, within a repository at the National Energy Research Scientific Computing Center (NERSC).

We ran our unWISE coaddition code to generate full-depth coadds uniformly incorporating all W1 and W2 exposures from 2010 January to 2020 December. No major code changes were made. However, we did patch the unWISE coadd software to account for the fact that, near the north ecliptic pole, the per-band integer frame coverage now surpasses $2^{15}$ = 32,768. The mean integer coverage over the whole sky is 279.6 (277.5) frames in W1 (W2). The minimum integer coverage is 106 (95) frames in W1 (W2). The maximum integer coverage is 33,032 (32,990) frames in W1 (W2), near the north ecliptic pole.

\begin{figure}[h!]
\begin{center}
\includegraphics[scale=0.8,angle=0]{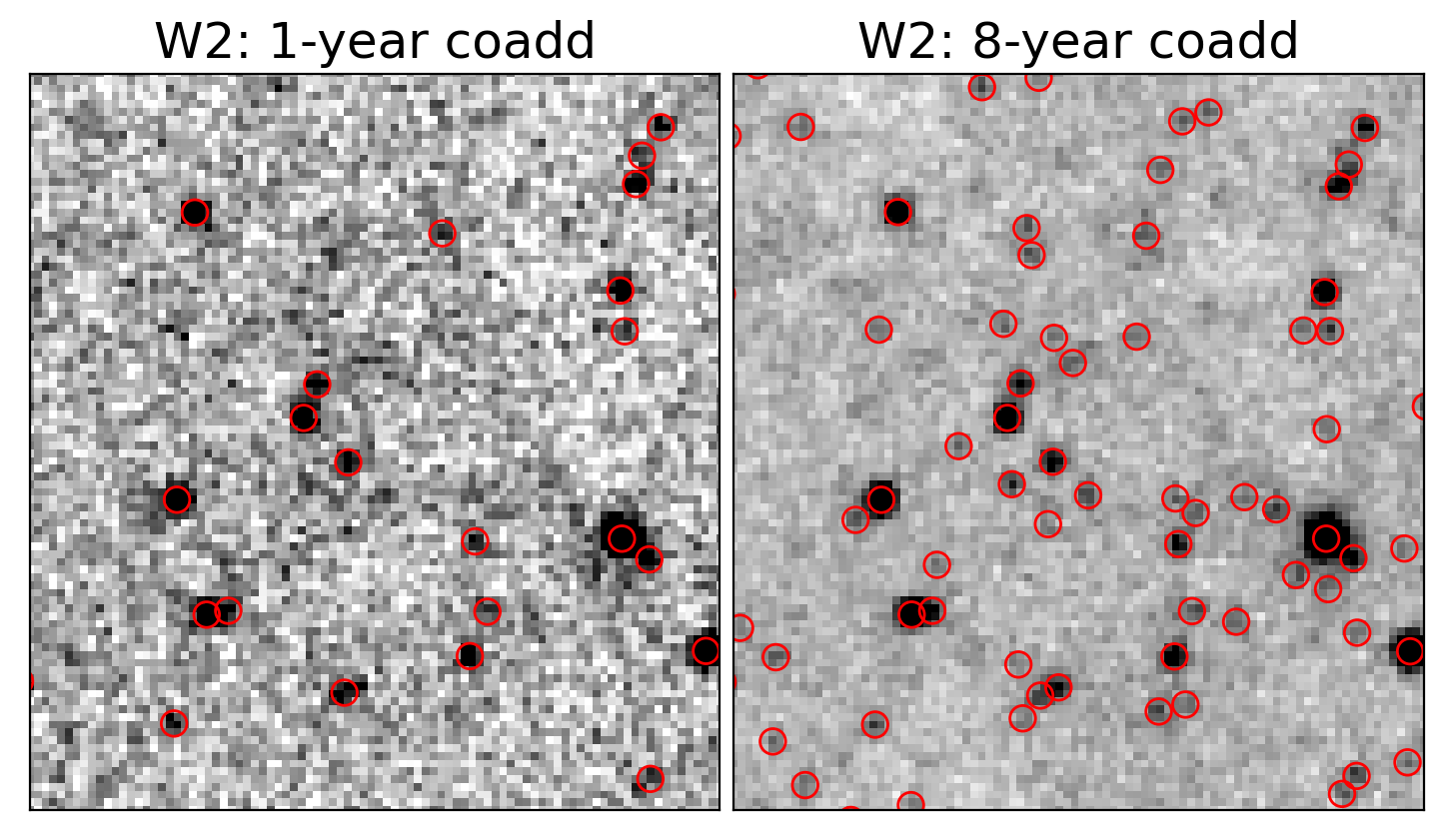}
\caption{A 4.1$'$ $\times$ 4.4$'$ region of sky at high Galactic latitude ($b_{gal} \approx 41.6^{\circ}$) centered on ($\alpha$, $\delta$) $\approx$ (150.1106$^{\circ}$, 1.3678$^{\circ}$). At left is a W2 coadd combining all pre-hibernation \textit{WISE} imaging, and at right is a W2 coadd that incorporates all available \textit{WISE} and NEOWISE data ranging from early 2010 through late 2020. The coadd at right benefits from $\sim$8$\times$ more total exposure time than the one at left. The grayscale stretch is the same in both cases. Red circles label the locations of $\ge 5\sigma$ sources detected using the \texttt{crowdsource} cataloging software \citep{unwise_catalog,decaps}. Averaging over the $\sim$2.4 square degree unWISE tile from which these cutouts have been extracted (\texttt{coadd\_id} = 1497p015), the 8-year (1-year) source density is 4.1 (1.8) per square arcminute, a factor of 2.3$\times$ increase. \label{fig:cosmos}}
\end{center}
\end{figure}

Figure \ref{fig:cosmos} shows a small patch of our 8-year W2 full-sky map, extracted from nearby the COSMOS region. One can perceive by eye that the depth is greatly enhanced through the inclusion of eight years of \textit{WISE} and NEOWISE imaging. The full-sky coadds described in this work can be downloaded from \url{https://portal.nersc.gov/project/cosmo/data/unwise/neo7/unwise-coadds/fulldepth/}.

\section{Acknowledgments}

This publication makes use of data products from the Wide-field Infrared Survey Explorer, which is a joint project of the University of California, Los Angeles, and the Jet Propulsion Laboratory/California Institute of Technology, funded by the National Aeronautics and Space Administration. This publication makes use of data products from the Near-Earth Object Wide-field Infrared Survey Explorer (NEOWISE), which is a joint project of the Jet Propulsion Laboratory/California Institute of Technology and the University of Arizona. NEOWISE is funded by the National Aeronautics and Space Administration.

This research is supported by the Director, Office of Science, Office of High Energy Physics of the U.S. Department of Energy under Contract No. DE-AC02-05CH1123, and by the National Energy Research Scientific Computing Center, a DOE Office of Science User Facility under the same contract. Additional support for DESI is provided by the U.S. National Science Foundation, Division of Astronomical Sciences under Contract No. AST-0950945 to the NSF's National Optical-Infrared Astronomy Research Laboratory; the Science and Technologies Facilities Council of the United Kingdom; the Gordon and Betty Moore Foundation; the Heising-Simons Foundation; the French Alternative Energies and Atomic Energy Commission (CEA); the National Council of Science and Technology of Mexico; the Ministry of Economy of Spain, and by the DESI Member Institutions. The authors are honored to be permitted to conduct astronomical research on Iolkam Du'ag (Kitt Peak), a mountain with particular significance to the Tohono O'odham Nation.

\bibliography{sample63}{}
\bibliographystyle{aasjournal}

\end{document}